\documentclass[lettersize,journal]{IEEEtran}

\usepackage{changes} % [final] 在完成后可隐藏标记
\definechangesauthor[name={Lian}, color=blue]{Lian}
\usepackage{amsmath,amsfonts}
\usepackage{algorithmic}
\usepackage{algorithm}
\usepackage{array}
\usepackage[caption=false,font=normalsize,labelfont=sf,textfont=sf]{subfig}
\usepackage{textcomp}
\usepackage{stfloats}
\usepackage{url}
\usepackage{verbatim}
\usepackage{graphicx}
\usepackage{cite}
\usepackage{hyperref}
\usepackage{booktabs} % 导入三线表需要的宏包
\usepackage{tabularray}
\usepackage{multirow}
\usepackage[normalem]{ulem}
\usepackage{xcolor}
\usepackage{ulem}
\usepackage{makecell}
\usepackage{booktabs}
\usepackage{tabularx}
\usepackage{amsmath}
\usepackage{url}
\usepackage{hyperref}
\usepackage{xr-hyper}          % 放在 hyperref 之后更好
\newcommand{\toolName}[1]{\textit{VisTCP}}
\newcommand{\revise}[1]{\textcolor{black}{#1}}

\begin{document}
%% Paper title.
\title{\toolName{}: A Visualization Framework to Construct 
%\added[id=Lian]{Knowledge-Graph-Based} Representation for Traditional Chinese Painting}
Knowledge-Graph-Based Representation for Traditional Chinese Painting}

\author{Zhiguang Zhou, Fengling Zheng, Miaoxin Hu, Lina You, Jin Wen, Huan Liu, Wei Zhang, \\Dekun Qian, Yuhua Liu, Wei Chen, Yigang Wang*, and Yong Wang*
        % <-this % stops a space
% \IEEEcompsocitemizethanks{Fengling Zheng and Zhiguang Zhou contribute equally to this work.}
% \IEEEcompsocitemizethanks{Zhiguang Zhou, Yuhua Liu, Yigang Wang, Jin Wen and Miaoxin Hu are with the School of Media and Design, Hangzhou Dianzi University (e-mail: zhgzhou@hdu.edu.cn, liuyuhua@hdu.edu.cn, yigang.wang@hdu.edu.cn, 231330023@hdu.edu.cn, miaoxinhu@outlook.com). Fengling Zheng and Lina You are with the School of Computer Science, Hangzhou Dianzi University (e-mail: fenglingzheng@hdu.edu.cn, linayou@hdu.edu.cn). Huan Liu, Wei Zhang and Wei Chen are with the State Key Lab of CAD\&CG, Zhejiang University (e-mail: alisalh@zju.edu.cn, zw\_yixian@zju.edu.cn, chenvis@zju.edu.cn). Yong Wang is with the College of Computing and Data Science, Nanyang Technological University (e-mail: yong-wang@ntu.edu.sg).
% \IEEEcompsocitemizethanks{Corresponding author: Yigang Wang and Yong Wang}
\IEEEcompsocitemizethanks{
  \IEEEcompsocthanksitem Equal contribution: Fengling Zheng and Zhiguang Zhou contributed equally to this work.
  \IEEEcompsocthanksitem {Zhiguang Zhou, Miaoxin Hu, Jin Wen, Dekun Qian, Yuhua Liu and Yigang Wang are with the School of Media and Design, Hangzhou Dianzi University (e-mail: zhgzhou@hdu.edu.cn, miaoxinhu@outlook.com, 231330023@hdu.edu.cn, qiandekun@hdu.edu.cn, liuyuhua@hdu.edu.cn, yigang.wang@hdu.edu.cn). Fengling Zheng and Lina You are with the School of Computer Science, Hangzhou Dianzi University (e-mail: fenglingzheng@hdu.edu.cn, linayou@hdu.edu.cn). Huan Liu, Wei Zhang and Wei Chen are with the State Key Lab of CAD\&CG, Zhejiang University (e-mail: alisalh@zju.edu.cn, zw\_yixian@zju.edu.cn, chenvis@zju.edu.cn). Yong Wang is with the College of Computing and Data Science, Nanyang Technological University (e-mail: yong-wang@ntu.edu.sg).}
  \IEEEcompsocthanksitem *Corresponding authors: Yigang Wang and Yong Wang}
% \wy{Pls double check if the format here is correct by comparing it with the original template. The correct command here should be % \IEEEcompsocitemizethanks{\IEEEcompsocthanksitem xxx} 
% }

}

% The paper headers
\markboth{IEEE TRANSACTIONS ON HUMAN-MACHINE SYSTEMS , VOL. xx, NO. x, JUNE 20xx}%
{Shell \MakeLowercase{\textit{et al.}}: A Sample Article Using IEEEtran.cls for IEEE Journals}

\maketitle
%% Author ORCID IDs should be specified using \authororcid like below inside
%% of the \author command. ORCID IDs can be registered at https://orcid.org/.
%% Include only the 16-digit dashed ID.
%\author{%
  %\authororcid{Josiah S.\ Carberry}{0000-0002-1825-0097},
  %Ed Grimley, and 
  %Martha Stewart
%}

%\authorfooter{
  %% insert punctuation at end of each item
  %\item
  	%Josiah Carberry is with Brown University.
  	%E-mail: jcarberry@example.com
  %\item
  	%Ed Grimley is with Grimley Widgets, Inc.
  	%E-mail: ed.grimley@example.com.

  %\item Martha Stewart is with Martha Stewart Enterprises at Microsoft
  %Research.
  	%E-mail: martha.stewart@example.com.
%}

%% Abstract section.
\begin{abstract}

Structured representation can characterize semantic objects and relationships in images. It provides a possible effective way for the semantic understanding of Traditional Chinese Paintings (TCPs) to better support archaeology and art history research.
However, most image-oriented structured representation methods perform poorly on TCPs, due to two major challenges: 1) the objects and events of TCPs
exhibit substantial differences from modern natural images, which results in semantic misunderstandings of TCPs; and  2) it is difficult to achieve accurate identification of ancient objects and events in TCPs, even for domain experts.   
In this paper, we propose \toolName{}, a visualization framework that combines a TCP-oriented intelligent model and expert knowledge, which enables art historians to achieve trustworthy structured representations of TCPs in a human-in-the-loop manner. Firstly, we conduct a pilot study with 
three
domain experts to build a semantic taxonomy of TCPs. Then,  expert-annotated data are used to train a TCP-oriented structured representation model, which can automatically extract meaningful objects and their relationships in TCPs. To inform users of the model uncertainty, we design a joint embedding visualization view to show the differences between expert annotations and model predictions. This allows users to refine the structured representation based on their domain knowledge,
enabling iterative optimization of the model. Finally, we conduct a case study, a usage scenario, and expert interviews on a real dataset to demonstrate the effectiveness of \toolName{} in supporting the structured representation and semantic understanding of TCPs.
% Finally, we conduct a case study, a usage scenario, and expert interviews using a real dataset to demonstrate the effectiveness of \toolName{} in the structured representation and semantic understanding of TCPs.

\end{abstract}

%% Keywords that describe your work. Will show as 'Index Terms' in journal
%% please capitalize first letter and insert punctuation after last keyword

%% A teaser figure can be included as follows

%% Uncomment below to disable the manuscript note
%\renewcommand{\manuscriptnotetxt}{}

%% Copyright space is enabled by default as required by guidelines.
%% It is disabled by the 'review' option or via the following command:
%\nocopyrightspace

%%%%%%%%%%%%%%%%%%%%%%%%%%%%%%%%%%%%%%%%%%%%%%%%%%%%%%%%%%%%%%%%
%%%%%%%%%%%%%%%%%%%%%% LOAD PACKAGES %%%%%%%%%%%%%%%%%%%%%%%%%%%
%%%%%%%%%%%%%%%%%%%%%%%%%%%%%%%%%%%%%%%%%%%%%%%%%%%%%%%%%%%%%%%%

%% Tell graphicx where to find files for figures when calling \includegraphics.
%% Note that due to the \DeclareGraphicsExtensions{} call it is no longer necessary
%% to provide the the path and extension of a graphics file:
%% \includegraphics{diamondrule} is completely sufficient.
\graphicspath{{figs/}{figures/}{pictures/}{images/}{./}} % where to search for the images

%% Only used in the template examples. You can remove these lines.
% \usepackage{tabu}                      % only used for the table example
% \usepackage{booktabs}                  % only used for the table example
% \usepackage{lipsum}                    % used to generate placeholder text
% \usepackage{mwe}                       % used to generate placeholder figures

% %% We encourage the use of mathptmx for consistent usage of times font
% %% throughout the proceedings. However, if you encounter conflicts
% %% with other math-related packages, you may want to disable it.
% \usepackage{mathptmx}                  % use matching math font

%%%%%%%%%%%%%%%%%%%%%%%%%%%%%%%%%%%%%%%%%%%%%%%%%%%%%%%%%%%%%%%%
%%%%%%%%%%%%%%%%%%%%%% START OF THE PAPER %%%%%%%%%%%%%%%%%%%%%%
%%%%%%%%%%%%%%%%%%%%%%%%%%%%%%%%%%%%%%%%%%%%%%%%%%%%%%%%%%%%%%%%

%% The ``\maketitle'' command must be the first command after the
%% ``\begin{document}'' command. It prepares and prints the title block.
%% the only exception to this rule is the \firstsection command

\section{Introduction}

\IEEEPARstart{W}{ith} %\added[id=Lian]{
% the digitalization of traditional Chinese painting, the use of computer technology to conduct research on ancient paintings has received widespread attention. From source exploration to authenticity identification, this involves various semantic related tasks, including semantic retrieval, question answering, and visual reasoning. 
 the digitization of traditional Chinese paintings(TCPs), computational methods have gained substantial attention in the fields of archaeology and art history, particularly for provenance analysis and authentication. However, art historians find that existing methods for structured understanding of TCPs, such as image classification, segmentation, and object detection, are inadequate for capturing the complex semantic content, which includes not only diverse objects, but also rich semantic events within the paintings. 
 These methods often fail to support advanced semantic tasks such as semantic retrieval, question answering, and visual reasoning due to their inability to effectively represent semantic information.
%}
 % However, art historians find it is far from ideal to conduct TCP analysis with the existing methods for structured understanding of TCPs such as image classification, image segmentation, and object detection, due to their fail of to represent the complex sematic content that not only have diversity objects but also rich semantic events within TCPs.
% These methods fail to support high-level semantic understanding and analysis such as semantic retrieval, question answering, and visual reasoning,due to their fail of semantic information representation.
% 随着中国传统绘画的数字化,借助计算机技术开展古画研究受到了广泛关注，从来源探究到真假鉴定，这涉及到各种语义相关任务，包括语义检索、问答以及视觉推理。

%\added[id=Lian]{
Knowledge graphs or scene graphs have been proposed for semantic representation of natural and chart images due to their ability to encode entities and relationships within images\cite{xu:2017:scene}.
%}
It has been proven capable of supporting various high-level tasks, such as structured knowledge understanding\cite{10387715}, semantic retrieval\cite{8100249}, and visual question answering \cite{8099827}. 
Inspired by this, it is highly desirable to characterize diverse objects of TCP and the inherent semantic relationships among them with a knowledge graph, thereby benefiting downstream tasks and empowering research in archaeology and art history.

% Inspired by this, it is highly desirable 
% to construct structured representations for TCPs
% to characterize diverse objects of a TCP and the inherent semantic relationships among them with the structured representation, so that benefiting downstream tasks and empowering archaeology and art history research.

%\added[id=Lian]{Various automated methods}
% structured representation 
 Various automated methods have been developed based on large image datasets training, such as Visual Relationship Detection\cite{tajrobehkar2021align} and Scene Graph Generation (SGG) \cite{tang:2020:unbiased}, to alleviate difficulties in repetitive and time-consuming knowledge extraction. However, 
 %\added[id=Lian]{
 in contrast to the large-scale data volume of natural images and the relatively simple visual elements and relationships in chart images, the sparse data volume, complexity of semantic content, and the diversity and professionalism of artistic styles in TCPs present significant challenges for knowledge graph construction.
 %}
% However, these methods are mainly developed for natural images and perform poorly in constructing structured representations for TCPs. The major challenges come from two aspects.
Specifically, \textbf{the objects and events depicted in TCPs differ markedly from those in natural images.
% , with substantial variations in both semantic content and image characteristics.
}
% While existing automatic methods can help mitigate some of the challenges in knowledge extraction, they cannot be directly applied to TCPs due to the lack of effective labels and sufficient sample sets for model training.
% For instance, we can find that there are some misidentifications and elements that have not been detected, as shown in Figure \ref{fig1-1}. Those objects that belong specifically to ancient China, such as the wish-granting sceptre, painted screen, horsetail whisk and wine vessel, are hard to be detected by machine learning models (Figure \ref{fig1-1}). 
% 一、古画在语义内容上以及图像特征表现与自然图像的差异表现给古画的自动知识抽取带来挑战。
% 二、艺术风格的差异导致不同的东西会有很不同的特征表现
Moreover, 
\textbf{
the accurate knowledge extraction is challenging due to significant variations in artistic styles across different TCPs.}
% For instance, some ancient objects like the wish-granting sceptre (Figure \ref{fig1-1}(B)) are hard to be recognized even by domain experts. In addition, the object labeled as a fan was erroneously identified as a modern racket in Figure \ref{fig1-1}(A), the chess board was misidentified as a box, and the ancient man was erroneously identified as a woman in Figure \ref{fig1-1}(B).   
% It can be seen that the resulting uncertainty in element identification impedes the acquisition of reliable knowledge graph representations for TCPs.
% The resulting uncertainty associated with element identification prevents obtaining credible structured representations for TCPs.

%\added[id=Lian]{
In this work, we aim to fill this gap by developing an interactive visualization framework to help users construct knowledge graph representations of TCPs. To meet comprehensive semantic representations of knowledge-graph-based structure for TCPs, we 
% acquire TCP's knowledge-graph-based structure for comprehensive semantic representation, we 
conducted an expert study with two art historians and analyzed insights from multiple rounds of expert interviews. Guided by the obtained high-level knowledge-graph-based representation structure, we designed and implemented \toolName{}, an interactive visualization tool,
%}
% To overcome the above challenges, we propose \toolName{}, a visualization framework 
that enables users to autonomously extract knowledge while enhancing their engagement in the construction of the TCP knowledge graph representation.
% \begin{figure}[htb]
%     \centering
%     % \vspace{-2em}
%     \includegraphics[width=\linewidth]{figure2-1.png}
%     \caption{Analysis of structured representations of different TCPs generated by intelligent model.}
%     \label{fig1-1}
%      % \vspace{-2em}
% \end{figure}
% Firstly, we conduct an expert study to develop a semantic taxonomy of TCP labels and establish an initial set of expert-annotated samples. Based on this, a structured representation model tailored for TCPs is then built by transferring learning from the object detection model Mask R-CNN and the relationship inference model Total Direct Effect (TDE). 
We first conduct an expert study to develop a semantic taxonomy for TCP entities and relations and to establish an initial set of expert-annotated samples. Building upon this, a structured representation model tailored for TCPs is constructed through transfer learning based on the object detection framework Mask R-CNN and the relationship inference model Total Direct Effect (TDE).
Furthermore, we present a joint embedding visualization that aligns expert-annotated elements with model predictions, revealing uncertainty and incorporating expert knowledge for credible refinement of structured representations. In addition, feedback from art historians and newly annotated samples are collected for iterative model improvement, further reducing human effort in constructing structured representations of TCPs. Finally, \revise{a case study, a usage scenario,} and expert interviews demonstrate that \toolName{} effectively integrates intelligent models and prior expert knowledge to efficiently construct the structured representation for TCPs.
This work mainly makes the following contributions:
\begin{itemize}
  \item We conduct an expert study to derive a semantic taxonomy and construct an expert-annotated dataset for TCPs, upon which a TCP-oriented knowledge extraction model is developed to enable comprehensive semantic representations of TCPs.
  % We derive a semantic taxonomy for the TCP knowledge graph structure to enable comprehensive semantic representation, along with an expert-annotated dataset, collectively supporting the construction of a TCP-oriented knowledge extraction model through expert study.  
 \item We develop a visualization framework, \toolName{}, that tightly integrates 
joint embedding visualization with human--AI collaborative interaction, 
enabling art historians to refine structured representations and further 
optimize the intelligent model for structured representation.
 % We develop a visualization framework, \toolName{}, that tightly integrates a joint embedding visualization method with human-AI collaborative interactions, enabling art historians to refine structured representations and optimize the intelligent structured representation model.  
  \item We conduct evaluations through a case study, a usage scenario, and expert interviews, demonstrating both the effectiveness of knowledge-graph-based representations in advancing TCP semantic understanding and the usefulness of \toolName{}.

\end{itemize}

\section{Related Work}
% This section reviews related studies from two complementary perspectives: 
% structured representation for Traditional Chinese Painting understanding, 
% and interactive visualization for human–AI collaboration.
\subsection{Structure Representation for Traditional Chinese Painting understanding}
Understanding Traditional Chinese Painting (TCP) requires interpreting multiple intertwined components, pictorial scenes, inscriptions, seals, and overall aesthetic composition. 
The pictorial scene, comprising \emph{figures, objects, and landscapes} with their associated \emph{events and spatial relationships}, constructs a narrative and symbolic world that reflects key dimensions of Chinese history and culture, forming the basis for the comprehensive understanding of TCPs~\cite{mccausland2013telling}.

Existing computational studies on TCP have explored various forms of \emph{structured representation} to enhance digital understanding, supporting tasks such as content-based retrieval, classification, and visual exhibition. 
Early works extracted handcrafted visual features describing texture, layout, and composition for content-based retrieval~\cite{zhang:2004:modelling,hung：2018：study}. 
With the advent of deep learning, CNN-based methods were applied to painting classification and element detection~\cite{10.1007/978-3-030-00009-7_22,meng:2019:elements,JIANG2019280}, yielding higher-level representations of visual structure. 
Interactive visualization systems, such as ScrollTimes for element validation~\cite{DBLP:journals/corr/abs-2306-08834} and audio-visual annotation frameworks for immersive presentation~\cite{7726076,10.1145/2307723.2307725}, further integrated visual and contextual information to improve interpretability. 
However, existing approaches mainly capture perceptual- and object-level structures, while the semantic and relational layers that embody contextual and symbolic meanings remain largely unexplored. 
Consequently, the absence of higher-level semantic representations hinders comprehensive understanding of TCPs.

Knowledge graphs (KGs) provide an interpretable framework for organizing entities, attributes, and relationships into an interconnected semantic network, enabling knowledge discovery and contextual inference~\cite{10.1007/s11263-016-0981-7,DBLP:journals/corr/abs-2307-14227}. 
Adopting a knowledge-graph-based representation enables the structured encoding of symbolic and hierarchical relations in TCPs, supporting semantic understanding beyond conventional image-level analysis.
However, constructing such representations demands domain expertise to accurately interpret the cultural semantics embedded in paintings. To this end, we construct a domain semantic taxonomy and an expert-annotated TCP knowledge graph that form the basis for expert-in-the-loop visual analysis and knowledge extraction.

\subsection{Interactive Visualization for Human–AI collaboration}
Interactive visualization has become an effective paradigm for integrating human expertise with machine intelligence. 
Through interactive exploration, it enables users to interpret and refine computational results, thereby bridging data-driven models with human reasoning. 
Such integration has demonstrated improvements in model reliability, interpretability, and data quality.
% \cite{9903604}
Building on these advances, visualization research can be systematically categorized into \emph{model-oriented} and \emph{data-oriented} approaches.
Model-oriented methods focus on visualizing and refining model behavior through interpretable representations.
For instance, RetainVis~\cite{8440842} and ProtoSteer~\cite{8827944} allow users to incorporate domain knowledge into deep sequence models, 
while DRAVA~\cite{10.1145/3544548.3581127} aligns semantic latent dimensions with human concepts via interactive manipulation. 
Data-oriented systems, on the other hand, focus on instance-level feedback and active learning. 
IRVINE~\cite{9552903} integrates interactive clustering and labeling for acoustic analysis, 
and RCLens~\cite{7939996} leverages user feedback to iteratively identify rare categories and improve labeling precision.

These studies collectively highlight that visualization not only supports model understanding but also enables interactive knowledge construction through uncertainty reasoning and active learning. 
Building upon these insights, our VisTCP framework employs uncertainty-aware visualization and expert feedback loops 
to facilitate human–AI collaboration in constructing and refining knowledge-graph representations of Traditional Chinese Paintings.

\section{Requirement Analysis and Framework Overview}
This section outlines the design requirements obtained through close cooperation with domain experts and provides an overview of the proposed visualization framework.

\begin{figure*}[tb]
    \centering
    \includegraphics[width=\linewidth]{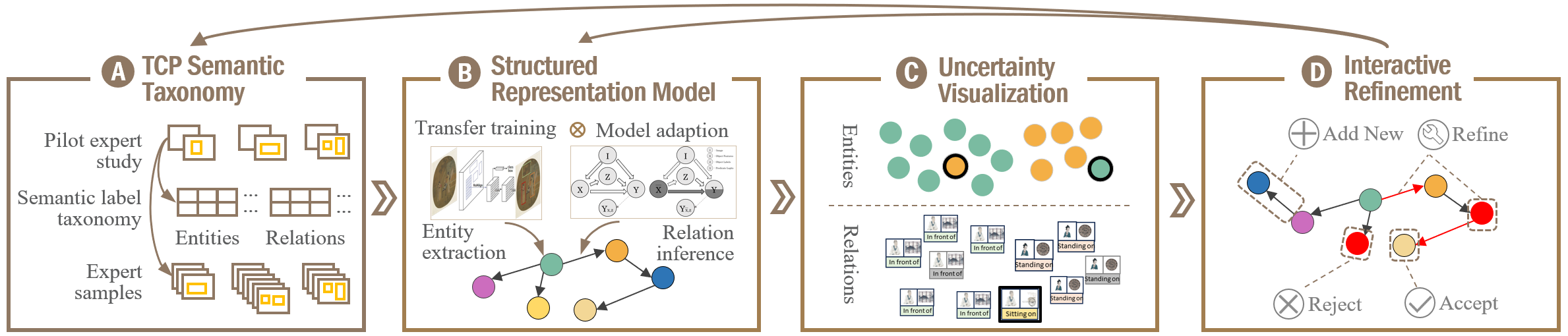}
    \caption{The framework architecture of \toolName{} contains four modules: a TCP semantic taxonomy module (A), a TCP-oriented model structured representation module (B), a visualization module (C), and an interactive refinement module (D).
    }
    \label{fig3}
\end{figure*}

\subsection{Requirement Analysis}
This study involved two stages and included three art historians working on TCPs and one expert experienced in the intersection of art imagery and computer vision, focusing on the construction and management of a digital TCP database. We collaborated with these four experts over two years. In the first stage, the art historians showed significant interest in developing a structured representation of TCPs for its powerful capabilities in supporting semantic retrieval and intelligent question-answering. We engaged with the experts' daily research activities for three months, encouraging them to describe any challenges they encountered in a think-aloud manner. Based on their feedback, we formed initial task requirements. Through multiple iterations of proposing solutions and receiving expert feedback, we ultimately defined the following requirements to empower art historians with enhanced knowledge interpretation capabilities, addressing their challenges in repetitive and uncertain knowledge extraction for the structured representation construction of TCPs.
% {Through close collaboration with the three domain experts over the past two years, we summarized the requirements to empower art historians with enhanced knowledge understanding capabilities, geared toward their challenges in repetitive and uncertain knowledge extraction for structured representation construction of TCP.}

\textbf{R1. Build a structured representation model for TCP.}
% Existing automatic structured representation methods have effectively reduced repetitive and time-consuming knowledge extraction. However, the substantial differences between ancient and modern objects and events inevitably lead to misunderstanding. Experts indicate that it is necessary to build a TCP-oriented structured representation model, to support a comprehensive understanding and ease the difficulty of knowledge extraction.  

\textit{R1.1 Construct TCP semantic taxonomy.}
% \textbf{R1.1 Construct TCP semantic taxonomy.}
To enrich semantic comprehension in Chinese visual culture, numerous comprehensive classification systems, such as Iconclass and CIT \cite{CIT}, have been advanced for visual content description.  While these established frameworks encompass a wide range of objects in Chinese visual culture, they overlook crucial semantic representations pertaining to ancient Chinese events. Consequently, experts advocate developing a specialized semantic taxonomy tailored for Traditional Chinese Paintings (TCPs) to foster semantic understanding of these cultural artifacts.

\textit{R1.2 Build structured representation model.}
% \textbf{R1.2 Build Structured representation model.} 
% Constructing structured representations is a repetitive and time-consuming knowledge extraction process. The experts indicated the importance of building a structured representation model following the TCP semantic taxonomy, to generate a high-quality initial structured representation, achieve acceleration, and reduce human effort in knowledge extraction.
Constructing structured representations is a repetitive and time-consuming process. Experts emphasized the need for a structured representation model aligned with the TCP semantic taxonomy to generate high-quality initial representations, accelerate knowledge extraction, and reduce human effort.

\textbf{R2. Prompt uncertainty with prior expert knowledge for structured representation refinement.} 
% Interpreting TCP is riddled with many uncertainties due to its historical attributes. Experts have illustrated that even experienced scholars must meticulously cross-reference their comprehension with precise knowledge. This practice also becomes indispensable when employing intelligent models. Experts emphasize the imperative need to incorporate prior expert knowledge to evoke uncertainty and refine structured representation results. 

\textit{R2.1 Visualize the uncertainty of knowledge extraction.}
Searching for prior references is a fundamental approach to attaining a precise comprehension of entities and semantic relationships in TCPs. While incorporating intelligent models can mitigate hurdles in knowledge extraction, both experts advocate for incorporating intuitive visual indicators alongside expert-driven prioritized knowledge to augment comprehension of extraction accuracy and signal uncertainties.

\textit{R2.2 Refine interactive structured representation.}
% \textbf{R2.2 Refine interactive structured representation.}
The consensus among experts underscores the necessity of integrating prior expert knowledge and user-friendly interactions to facilitate users in iteratively refining the initial structured representation of TCP. Additionally, experts highlight the need for intelligent models to efficiently adapt to the distinct visual characteristics of TCPs through user feedback. This adaptability is crucial for the accurate and efficient construction of the initial structured representations of TCPs.

\subsection{Framework Overview}
The aforementioned requirements motivated us to develop a visualization framework, \toolName{}, that integrates intelligent models and prior expert knowledge into an intuitive visual interface, empowering art historians to construct a robustly structured representation for the semantic understanding of TCPs.
% The aforementioned requirements motivated us to develop a visualization framework integrating intelligent models and prior expert knowledge in an intuitive visual interface, to empower art historians in constructing a robustly structured representation for semantic understanding of TCPs.  
% interpreting TCP with greater ease and precision. 
\toolName{} (Figure \ref{fig3}) includes four parts. First, a study with domain experts is conducted to create a TCP semantic taxonomy and an expert-annotated sample set (Figure \ref{fig3}(A))
% , which are then utilized 
for TCP-oriented structured representation model building
(Figure \ref{fig3}(B)). Subsequently, we propose an uncertainty visualization that jointly maps semantic labels and image features from both model predictions and expert-annotated samples, to highlight uncertain entities and relationships
% Afterward, the uncertainty visualization is proposed to map both semantic labels and image features of model prediction and expert-annotated sample set, for uncertainty entities or relationships prompt
% prompting uncertainty entities or relationships
% triplet events 
% for correction 
(Figure \ref{fig3}(C)). Finally, the semantic label taxonomy undergoes iterative refinement through users' interactive inspection and updating of these entities or relationships (Figure \ref{fig3}(D)). Additionally, the model is iteratively fine-tuned based on user feedback to generate higher-quality structured representations. Through continual iterative optimization, the structured representation of TCP can be achieved more swiftly and accurately, with minimized manual intervention.

% \begin{figure*}[tb]
%     \centering
%     \includegraphics[width=\linewidth]{Figure2.png}
%     \caption{The pipeline of the proposed visual system.}
%     \label{fig3}
% \end{figure*}

\section{Structure Representation Model}
% We propose a visualization framework, \toolName{}, that integrates automatic knowledge extraction and interactive refinement to build structured representations of TCPs.
% This section introduces three computational methods of the proposed framework: knowledge extraction model setup, visual-semantic joint embedding, as well as model fine-tuning.
% This section constructs the structured representation model for TCPs by: (i) deriving a semantic taxonomy and curating an expert-annotated dataset for TCPs via an expert study (see Appendix A for the full description)
This section presents the construction of the structured representation model for TCPs, which involves (i) deriving a semantic taxonomy and constructing an expert-annotated dataset through an expert study (see Appendix~A for the full description), (ii) instantiating an SGG-based model with Mask-RCNN for object detection and TDE for unbiased relation inference, and (iii) integrating an active-learning loop for iterative refinement.

\subsection{Semantic Taxonomy and Expert-Annotated Dataset Construction}
\label{sec:coretax}
% We adopt a compact taxonomy with four entity classes (\emph{human, natural scenery, animal, artifacts}) and two relation types (\emph{event, location}).
We adopt a compact taxonomy with four entity classes (\emph{human, natural scenery, animal, and artifact}) and two relation types (\emph{event} and \emph{location}).

\textbf{Derivation method.} The taxonomy was derived via an expert study with three TCP specialists using our annotation prototype: starting from a seed vocabulary compiled from museum thesauri and prior scholarship~\cite{CIT}, experts annotated candidate entities and relations in a pilot study of 500 paintings (300 figure, 100 landscape, 100 flower-and-bird paintings), were allowed to add/modify labels, and then reconciled disagreements in a consensus meeting. 

\textbf{Annotated dataset construction.} The same process produced a 500-image expert-annotated set that instantiates the taxonomy with object bounding boxes and relation triplets. Annotations were cross-checked across experts, conflicts were adjudicated, and each object/relation category was ensured to have at least 10 examples to support training and evaluation.

This set is used (i) to define detection/relationship categories for our Mask-RCNN~\cite{8237584} + TDE~\cite{tang:2020:unbiased} pipeline and (ii) to drive active learning for iterative model refinement. Detailed study design, full label inventories, and per-type mappings are provided in Appendix A ($Tables \uppercase\expandafter{\romannumeral 1}-\uppercase\expandafter{\romannumeral 2}$).

\begin{figure*}
\centering
\begin{minipage}{1\textwidth}
    \centerline{\includegraphics[width=1\textwidth]{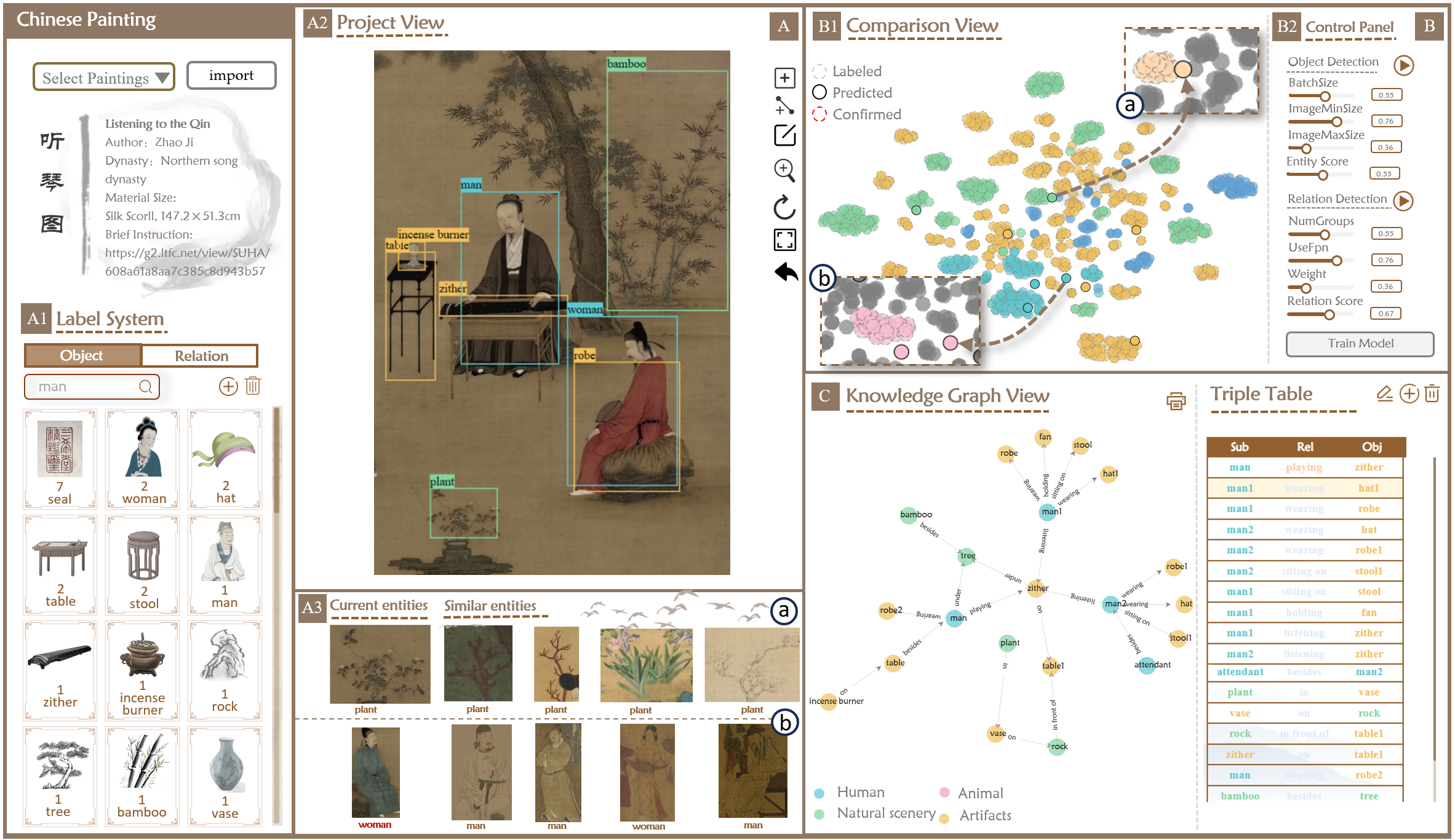}}
\end{minipage}
\caption{
 The visualization interface of \toolName{} consists of three components, including (A) an interactive refinement component provides the knowledge extraction results of a painting predicted by models and referenced samples for interactive refinement, (B) the uncertainty visualization component aligns the visual and semantic features of labeled and predicted elements to highlight ambiguities requiring expert verification, and (C) the structured representation component presents the obtained semantic graph from a TCP.}
\label{fig:interface}
\end{figure*}

\subsection{Structured Representation Model Setup}
The proposed structured representation model (R2) is built upon SGG\cite{tang:2020:unbiased}, a widely used model for the structured representation of natural images, and consists of two parts: 
the advanced object detector Mask-RCNN \cite{8237584} and unbiased relation inference method Total Direct Effect (TDE).  

\textbf{Object Detection.}
Mask-RCNN is a state-of-the-art deep learning algorithm for object detection, which is widely used for its ease of use and customization, accurate object detection, and faster processing. We adopt ResNet-101 as the backbone network architecture and the proposed TCP semantic taxonomy as the class labels, to build the object detection model for TCPs, to accommodate the characteristics of ancient China. 

\textbf{Relationship Inference.}
% Total Direct Effect Relation Inference designed for unbiased scene graph generation, is utilized to address the serious bias issue in SGG, which successfully achieves rich relationship prediction output.
The Total Direct Effect (TDE) relation inference model, designed for unbiased scene graph generation, is utilized to mitigate the bias problem in SGG, achieving richer and more diverse relationship predictions. We also employ ResNet-101 as the foundational network architecture for the prediction of relationships in TCPs, while the semantic relationship labels of TCP taxonomy are utilized to enable the model to adapt to TCPs. 
 
 \textbf{Loss Function.} TTo supervise transfer learning in the network, we use the cross-entropy loss to train the structured representation model for TCPs, which measures the distance between the ground truth $y$ and the predicted probability distribution $p$,
\begin{equation}
 L_{CE} =-\frac{1}{N} {\textstyle \sum_{i=1}^{N} {\textstyle \sum_{c=1}^{C}}y_{i,c}\log_{}{(p_{i,c})} }   
\end{equation}
$N$ is the number of samples, $C$ is the number of classes,  $y_{i,c}$ represents the ground truth label of class 
$c$ for sample $i$ ($1$ if the sample belongs to class $c$, $0$ otherwise), $p_{i,c}$ is the predicted probability of class $c$ for sample $i$.  
 
% The expert-annotated sample set in the pilot expert study is employed for training,  
% the experiment results confirm the trained model is well adapted to TCPs, as demonstrated in $Tables \uppercase\expandafter{\romannumeral 3}-\uppercase\expandafter{\romannumeral 4}$ of the Appendix B.
% These two components form the structured representation model for TCPs.
The expert-annotated sample set from the pilot expert study is used to train the model, and the experimental results confirm that the trained model is well adapted to TCPs, as demonstrated in Tables~\uppercase\expandafter{\romannumeral3}–\uppercase\expandafter{\romannumeral4} in Appendix~B. Together, these two components form the structured representation model for TCPs.

\subsection{Model Optimization}
We refine our TCP-oriented knowledge extraction model through active learning (R4), 
based on the collections of user feedback for uncertainties and an increasing number of annotated samples. 
The loss function for optimizing the model is set the same as model transfer learning, to ensure iterative improvement.
The important aspect of supporting model optimization is the selection of training samples. 
Here, we use annotated samples with joint embedding priority recommendations for model fine-tuning, which has been proven to be an effective active learning optimization strategy \cite{REAL}.

% \begin{figure*}[tb]
%     \centering
% \includegraphics[width=\linewidth]{figs/Figure4-1.png}
%     \caption{
%     % Example Scenario 2. 
%     The visualization interface of \toolName{} consists of three components. (A) The interactive refinement component provides the knowledge extraction results predicted by models and expert-annotated samples for interactive refinement. (B) The uncertainty visualization component presents the visual and semantic features between expert-annotated and predicted elements to prompt uncertainties that need to be checked. (C) The structured representation component presents the obtained semantic graph.}
%     \vspace{-3mm}
%     \label{fig:interface}
% \end{figure*}

\section{VisTCP Interface}
In this section, we introduce the visualization and interaction designs in \toolName{}.
% for presenting the structured representation identification results with explanations and to support trustworthy iterative refinement. 
The interface of \toolName{} is composed of three components, including Interactive Refinement Component (Fig. \ref{fig:interface}(A)), Uncertainty Visualization Component (Fig. \ref{fig:interface}(B)), and Structured Representation Component (Fig. \ref{fig:interface}(C)). 

\subsection{Uncertainty Visualization Component}
The Uncertainty Visualization Component
% Knowledge Extraction and Understanding component 
integrates a model panel
% to facilitate art historians
to achieve knowledge extraction, and incorporates a Comparison View to enhance comprehension of the extracted knowledge. 

% \subsubsection{Model Panel}
% \textbf{Support automatic knowledge extraction.} As shown in Fig. \ref{fig:interface}(B2), two models of knowledge extraction are supported here: 1) entity extraction, which detects objects and outputs corresponding semantic labels and bounding boxes; and 2) relation inference, which predicts semantic relations between object pairs. The initial structured representation is then integrated into the workflow.

% \subsubsection{Comparison View}
% It presents a joint visualization of embedding vectors for model predictions and expert-annotated samples (Fig.~\ref{fig:interface}(B1)), enabling art historians to intuitively assess results and pinpoint high-uncertainty elements for targeted validation.
\subsubsection{Model Panel}
\textbf{Support for automatic knowledge extraction.} 
As shown in Fig.~\ref{fig:interface}(B2), the system supports two knowledge extraction models: 
(1) \emph{Entity extraction}, detecting objects and outputting corresponding semantic labels and bounding boxes; and 
(2) \emph{Relation inference}, predicting semantic relations between object pairs. 
The resulting structured representation is then integrated into the workflow.

\subsubsection{Comparison View}
This view presents a joint visualization of embedding vectors for model predictions and expert-annotated samples (Fig.~\ref{fig:interface}(B1)), enabling art historians to intuitively assess results and pinpoint high-uncertainty elements for targeted validation.

\textbf{Visual–Semantic Joint Embedding.}
We jointly embed predicted and expert-annotated entities/relations (R3) using visual and semantic features to improve interpretability and surface uncertain items for rapid refinement.
Predicted elements are first paired with expert exemplars via text matching and feature similarity.
Object and relation features are extracted by our TCP-oriented model and mapped into a shared space where Euclidean distances are re-weighted by semantic labels, pulling same-label items together and pushing conflicting labels apart.
The resulting label-aware embedding underpins the Comparison View.

\begin{figure}[tb]
	\centering
	\includegraphics[width=3.5in]{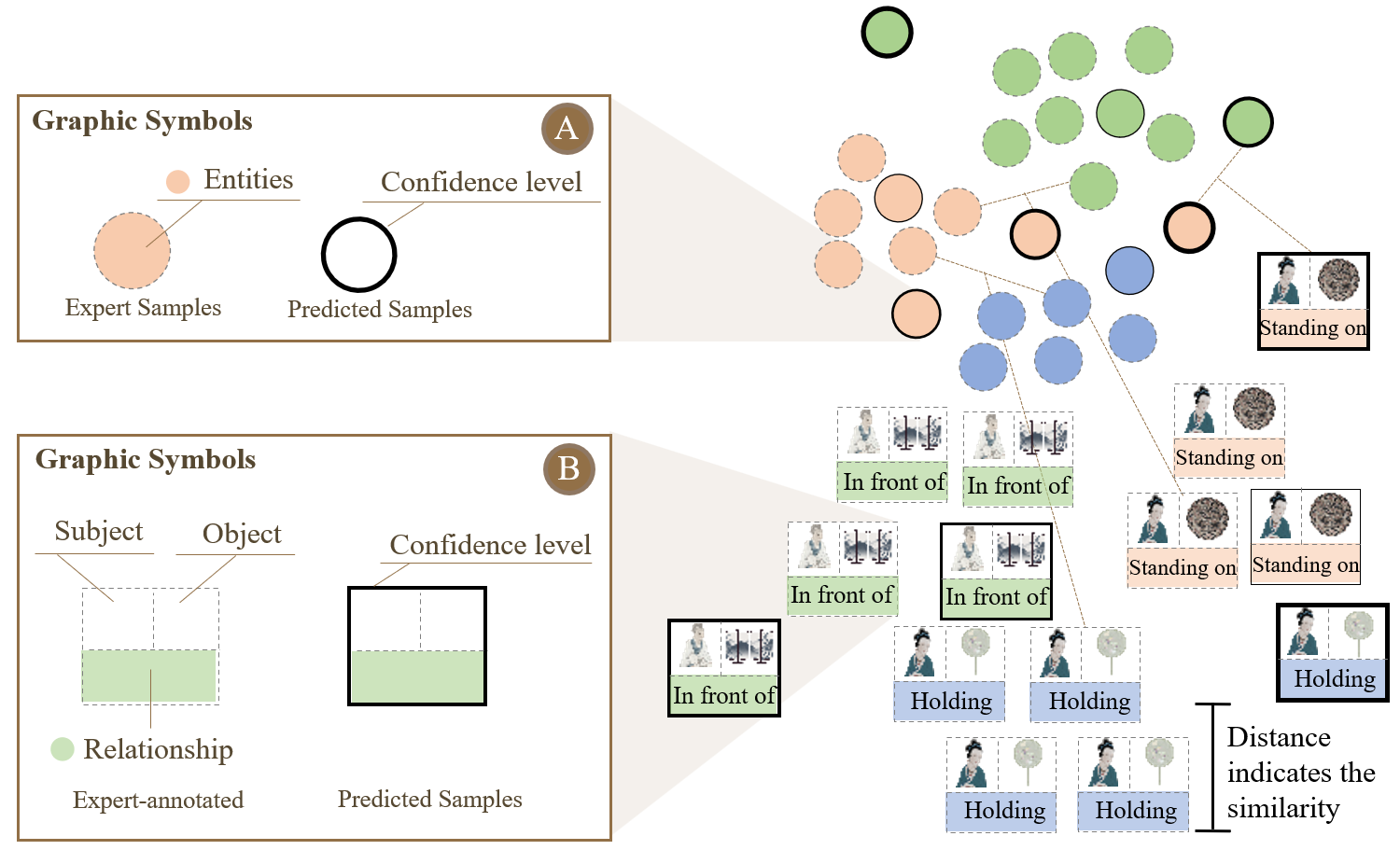}
	\caption{The joint embedding visualizations of the predicted and expert-annotated samples in the TCPs for objects (A) and semantic relationships (B).}
	% \vspace{-4mm}
 \label{fig4}
\end{figure}

\textbf{Encodings.}
We concurrently expose three facets: (1) feature-driven distances, (2) per-label neighborhoods, and (3) the alignment/misalignment of predictions vs.\ expert annotations.
To make these facets legible in one view, we project the embedding to 2D and encode aspects in distinct channels.
\emph{Entities} (Fig.~\ref{fig4}(A)) are shown as a scatterplot: point color encodes the semantic label; solid vs.\ dashed borders distinguish prediction vs.\ expert exemplar; and border thickness encodes prediction confidence.
\emph{Relations} (Fig.~\ref{fig4}(B)) use compact icons integrating subject–predicate–object; the same border scheme distinguishes sources.

\textbf{Layout by feature similarity.}
Embedded positions are governed by feature proximity.
We employ t-SNE (chosen for preserving local clusters) to maintain intra-class cohesion while separating inter-class neighborhoods, which facilitates side-by-side comparison of predictions and expert knowledge and highlights potential ambiguities.

\textbf{Uncertainty cues.}
Elements that lie far from their own label cluster, fall near competing-label clusters, or have low prediction confidence are flagged as high-uncertainty candidates and routed to the refinement workflow.

\subsection{Interactive Refinement Component}
The Interactive Refinement Component provides a robust visual coordination environment and extensive interactive functionalities to streamline the iterative refinement of knowledge extraction, guided by prior expert knowledge. The Label System View, Painting View, and Expert View collectively present an overview of initially identified results, the original painting image annotated with semantic labels, and recommended expert-annotated samples, fostering a coherent and efficient refinement process.

\subsubsection{Label System View}
% The Label System view 
It provides an overview of the proposed TCP semantic taxonomy (Fig. \ref{fig:interface}(A1)), accompanied by corresponding label icons, to clarify the distinctive characteristics of objects and events pertaining to ancient China. Entity or relation selection and accuracy adjustment are also provided in this overview for filtering, which allows users to focus on a certain class of interest for 
% knowledge extraction 
refinement.

\subsubsection{Painting View}
% The Painting view 
It presents the original painting image alongside the identified results
with bounding boxes for objects, lines for relationships, and the corresponding predictive labels (Fig. \ref{fig:interface}(A)). To enhance user efficiency in refining entities and relationships, while simultaneously reducing visual clutter resulting from numerous entities bounding boxes and relation lines, each entity's bounding box will be transformed into a precise point marker once the user confirms its semantic label. Similarly, after entity extraction, further optimization or augmentation of relationships will be conducted in accordance with the visual cues provided in the Comparison View (Fig. \ref{fig:interface}(B)). In instances involving entity pairs with relationships, these connections are initially depicted as dashed lines, which are later changed to solid lines upon confirmation.
% of the relationship.

\subsubsection{Expert View}
It displays expert-annotated samples that have the most similar characteristics to the selected predictive sample, for refinement reference (Fig. \ref{fig:interface}(A3)). The labels of expert-annotated samples can be transferred to predictive samples through a double-click action. 
Herein lies an in-depth exposition of expert annotation samples, encompassing specific annotations of entities or events alongside corresponding illustrations to furnish additional contextual information.

\textbf{Enabling interactive refinement.} Collaborating with all other views, six operations are supported in here: 1) Modify entity. Users can change the label of an entity by clicking the corresponding bounding box. 
2) Add entity. This action is initiated by a left-clicking arrow icon in the upper right corner, then draws a frame 
and selects a label to complete the entity add. The active input function is also provided for entity addition outside the label range to ensure the scalability of the label. 3) Delete entity. The user can remove the active bounding box corresponding to an entity via the \emph{Delete} button. 4) Modify relation. Users can delete, modify, or confirm the relationship label by double-clicking on the relationship line between two entities. 5) Add relation. The user can add the relationship by clicking on the first and last entities in sequence. 6) Zooming and panning tools are also provided to facilitate knowledge extraction and enhance the exploration of paintings.

\begin{figure*}
\centering
\begin{minipage}{1\textwidth}
    \centerline{\includegraphics[width=1\textwidth]{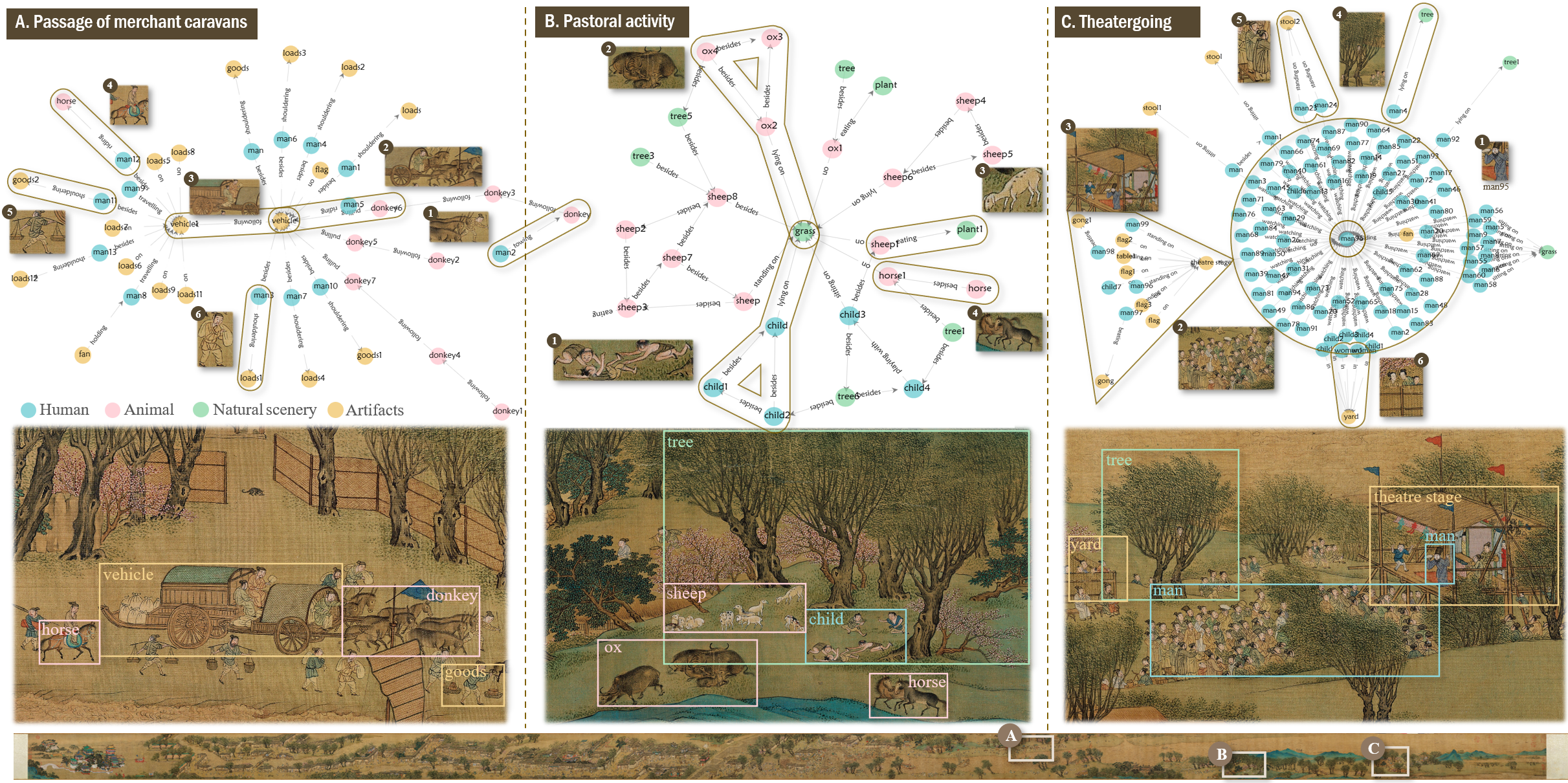}}
\end{minipage}
\caption{
 Structured representation facilitates the understanding of the famous Chinese painting ``Along the River During the Qingming Festival'' across three distinct scenes: the passage of merchant caravans (A), pastoral activity (B), and theatergoing (C).
}
\label{fig:representation}
\end{figure*}

\subsection{Structured Representation Component}

The Structured Representation Component (Fig. \ref{fig:interface}(C)) offers a precision depiction of the current painting, delineating distinct entity types through node colors (blue: Human; pink: Animal; green: Natural Scenery; orange: Artifact) and representing distinct relationships through line colors (grey: location; yellow: event). When performing entity or relation editing operations, the interface provides real-time updates to promptly reflect changes in the graph's status.
% When carrying out entity or relation editing operations, the view offers real-time updates, to reflect changes in the graph's status promptly.
Furthermore, it offers two distinct layout methods: force-oriented layout and custom layout, facilitating exploration and analysis of TCPs. In addition, the Triple Table records various interactions to collect user feedback for model improvement. 

% \begin{figure*}
% \centering
% \begin{minipage}{1\textwidth}
%     \centerline{\includegraphics[width=1\textwidth]{QingRiver.png}}
% \end{minipage}
% \caption{
%  Structured representation facilitates the understanding of the famous Chinese painting ``Along the River During the Qingming Festival'' across three distinct scenes: the passage of merchant caravans (A), pastoral activity (B), and Theatergoing (C).
% }
% \label{fig:representation}
% \end{figure*}

% \begin{figure*}
% \centering
% \begin{minipage}{1\textwidth}
%     \centerline{\includegraphics[width=1\textwidth]{Figure4-1.png}}
% \end{minipage}
% \caption{
%  The visualization interface of \toolName{} consists of three components, including (A) an interactive refinement component provides the knowledge extraction results of a painting predicted by models and referenced samples for interactive refinement, (B) the uncertainty visualization component visually compares the visual and semantic features between labeled and predicted elements to prompt ambiguities that needed to check, and (C) the structured representation component presents the obtained semantic graph from a TCP.}
% \label{fig:interface}
% \end{figure*}

\section{Evaluation}
%\added[id=Lian]{
This section evaluates the effectiveness of the knowledge graph for TCP comprehensive semantic representation, followed by a usage scenario and expert interviews with 6 art historians to demonstrate the applicability of \toolName{} in constructing structured representations.
%}
% We first conducted a case study
% to demonstrate the usefulness of the structured representation constructed using \toolName{}.
% Then, we presented a usage scenario and interviewed art historians to validate the effectiveness and usefulness of \toolName{} in constructing structured representations for TCPs.

\subsection{Case Study
}
\revise{In this case, we primarily showcase the structured representations of three different scenes depicted in \textit{Along the River During the Qingming Festival}: the passage of merchant caravan, pastoral activity, and theatrical spectacle, to illustrate the effectiveness of structured representations for TCP comprehensive semantic representation.} Fig.~\ref{fig:representation} 
% \replaced{To clarify the structured representation for semantic understanding, we illustrate it from a visualization perspective to facilitate information communication.}{}
presents the structured representations, which show significant differences aligned with their visual scenes, as follows: 

\textbf{Passage of merchant caravans.} Fig.~\ref{fig:representation}(A) illustrates the structured representation of a bustling caravan, which presents a uniform distribution of animal nodes (pink), human nodes (blue), and artifact nodes (orange). A close observation reveals that all the animals are donkeys, all the humans are men, and the artifacts mainly consist of loads and two vehicles. This setup highlights the participatory features of the elements in the scene. The elements form a chain-like layout based on primary semantic relationships, predominantly unfolding in the same direction. The chain starts with a man towing a donkey, followed by the donkey pulling a vehicle, the vehicle following another vehicle, and a man riding a horse following the vehicle. These events collectively depict the visual scene of the merchant caravan's passage. Furthermore, the presence of numerous men shouldering loads around the carts enhances the visual scene of the caravan's passage, contributing to a comprehensive semantic understanding of the scene.

% \textbf{Passage of merchant caravans.} Figure~\ref{fig:representation}(A) illustrates the structured representation of a bustling caravan, where we can find a uniform distribution of animal nodes (pink), human nodes (blue), and artifact nodes (orange). \revise{A close observation reveals that animals are all donkeys, humans are all men, while the artifacts most are loads and two vehicles.} This primarily expresses the participatory features of elements in the scene. In addition, they form a chain-like layout based on primary semantic relationships. The relationships in this chain-like structure primarily unfold in the same direction,  
% starting with a man towing a donkey, followed by descriptions of the donkey pulling a vehicle, the vehicle following another vehicle, and a man riding a horse following the vehicle, among other semantic events. These events collectively depict the visual scene of the passage of merchant caravans. Additionally, the presence of numerous men shouldering loads around the carts further reinforces the visual scene of the caravan's passage, contributing to the semantic understanding of the scene.

\textbf{Pastoral activity.} As illustrated in Fig. \ref{fig:representation}(B), the structured representation of a pastoral activity, 
primarily characterized by animal nodes (pink), natural scenery nodes (green), and human nodes (blue), thereby revealing a pastoral scene distanced from human settlements. A meticulous examination of this region reveals a plethora of structured representations, encompassing numerous boys and animals such as ox, sheep, and horses, engaged in various activities including playful interactions among the boys and animals eating grass. Conversely, the presence of artificial artifacts is notably sparse. 
% Consequently, 
% This 
% % structure
% vividly portrays the pastoral setting where boys are tending to livestock in the countryside.
This vividly portrays the semantic events in a pastoral setting, where boys are tending livestock in the countryside.
 
\textbf{Theatergoing.}
The structured representation (Fig. \ref{fig:representation}(C)) of theatrical spectacle exhibits a unique aggregated structure, primarily centered around the semantic event of individuals watching a person perform on a theatre stage.  
Additionally, the ``man95'' node, representing the central figure, is connected to another local structure.  Upon closer examination, it becomes apparent that this primarily describes individuals positioned on the same stage, involving events such as a man playing a drum, a table on the stage, and a flag on the stage. Together, they comprehensively depict the scene on the stage. 
Furthermore, this structured representation vividly delineates various interesting events in the theatrical scene, such as a man watching the performance from a tree, a man standing on a stool to watch the performance, and a woman observing the performance in the courtyard. The triplet descriptions
% of these semantic events 
also suggest latent semantic events in the scene, facilitating a comprehensive semantic representation of the theatrical scene.

In summary, we observe that the structured representations of the three distinct visual scenes depicted in ``Along the River During the Qingming Festival'' exhibit markedly different characteristics, with the rich semantic events in each scene clearly delineated. Additionally, the visualization of these structured representations highlights less conspicuous semantic events within the visual scenes, thereby enhancing semantic understanding.
% \revise{In summary, we can observe that the structured representations of the three distinct visual scenes depicted in ``Along the River During the Qingming Festival'' exhibit markedly different characteristics, while the rich semantic events contained in each scene are clearly delineated. 
% Additionally, the visualization of structured representations can also highlight less conspicuous semantic events in visual scenes, thereby enhancing semantic understanding.}

\subsection{Usage Scenario}
%\deleted[id=Lian]{In this section,  we introduce a usage scenario to showcase the usefulness of \toolName{} in constructing structured representations.}  
% In this usage scenario, 
An art historian specializing in figure paintings and deeply interested in applying computer technology to analyze TCPs was invited to use \toolName{} to construct the structured representation for the well-known Chinese painting \textit{The Night Revels of Han Xizai.}
% An art historian who studies figure paintings, with a deep-seated interest in employing computer technology for the analysis of TCPs, was invited to use \toolName{} to construct the structured representation for the famous Chinese painting, \textit{The Night Revels of Han Xizai.}

\begin{figure*}[tb]
    \centering
    \includegraphics[width=\linewidth]{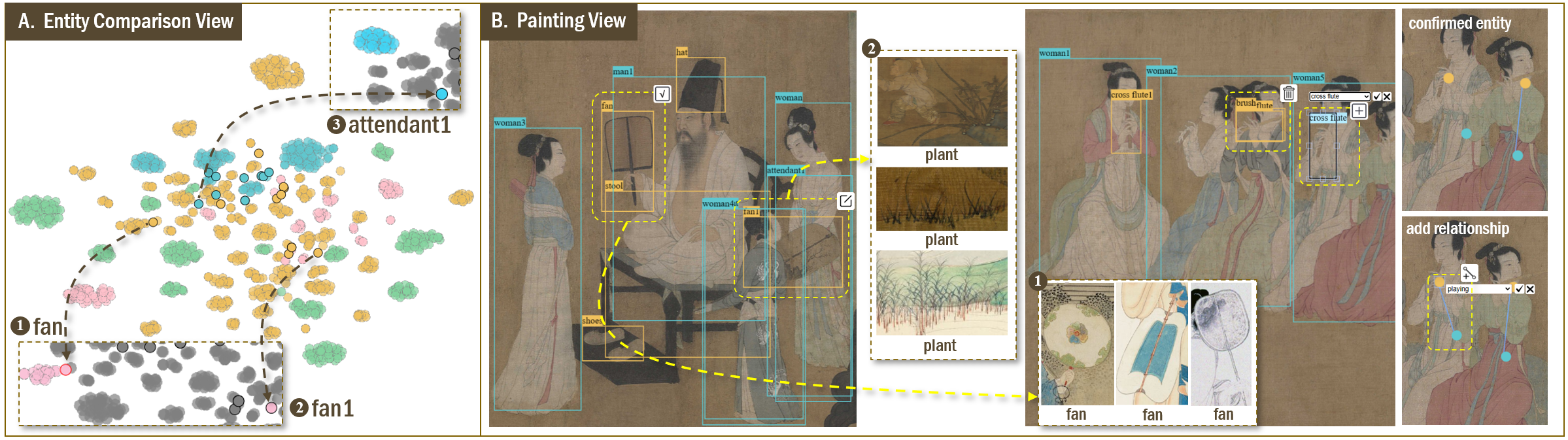}
    \caption{
    Constructing the structured representation for ``The Night Revels of Han Xizai''.
    The visualization of automatically generated structured representations supports understanding of result quality (A). Then, the art historian examined and refined uncertain elements to produce a credible structured representation (B).}
    \label{fig6}
\end{figure*}

\textbf{Understand the quality of 
automatic knowledge extraction.}
First, the art historian clicked the object detection button and examined the joint embedding visualization (Fig. \ref{fig6}(A)). He noticed that most predicted objects fell within the dense clusters of expert-annotated samples with the same class labels, indicating a high level of confidence in the detections. He then clicked on an orange predicted point and reviewed the expert samples in the expert view, confirming that its label was indeed a fan, consistent with the recommended label of the expert samples (Fig. \ref{fig6}(B-1)). The joint embedding visualization results garnered the art historian's trust. He then noticed some predicted points that were distant from the expert-annotated samples with the same class labels, suggesting a likelihood of incorrect identifications. The art historian concluded that the automatic object detection results required further refinement.
% First, the art historian clicked the object detection button and 
% looked at the joint embedding visualization (Figure \ref{fig6}(A)).
% % to compare the objects identified by the automatic knowledge extraction model and previously expert-annotated samples. 
% He noticed that most predicted objects fall in the dense clusters of expert-annotated samples with the same class label. This indicates that the detected objects have a heightened level of confidence. Then, he clicked one orange predicted point and reviewed the expert samples in the expert view, confirming its label was indeed a fan just like the recommended label of expert samples (Figure \ref{fig6}(B-1)). The outcomes of the joint embedding visualization have garnered the trust of the art historian. Then he noticed there were indeed some predicted points away from the expert-annotated samples with the same class labels. This indicates that it is highly likely that objects that have been incorrectly identified. 
% The art historian felt that the result of the automatic object detection required further refinement.

\revise{\textbf{Explore and refine elements with uncertainty.}}
The art historian reviewed the suspicious elements highlighted by the joint embedding visualization and found most were reasonable. For example, he noticed a blue point distant from its cluster (Fig. \ref{fig6}(A-3)) and, upon checking the expert view, saw that the object in the bounding box was incomplete, leading him to delete it. Exploring further, he observed two fans identified by the model in the Label System View. Clicking the icon, he found another fan far from the expert-annotated samples, suggesting misclassification. Examining the original painting, he found the fan identical to the previous one, with the deviation resulting from its plant-like illustrations (Fig.~\ref{fig6}(B-2)). Finally, he confirmed that the identified women’s labels aligned with a dense cluster of expert-annotated samples.
% The art historian inspected the suspicious elements prompted by the joint embedding visualization. He found that most of them are reasonable. For example, he noticed a detected blue point away from its corresponding dense cluster (Figure \ref{fig6}(A-3)), then he checked the expert view and found the object in the detected bounding box was incomplete, then he deleted it. To further explore the result, he noticed two fans were identified by the model in the Label System View, he clicked the icon to review and found another detected fan was substantially distant from the expert-annotated samples, suggesting a high prosbability of misclassification. By examining the original painting, he found that this fan was identical to the previous one. The deviation of the detected point from the expert samples was attributed to the presence of plant-like illustrations on the fan (Figure \ref{fig6}(B-2)).
% Then, he further checked the identified women, which all were in the dense cluster of expert-annotated samples with the same class label,
% and confirmed that the predicted label of the model was indeed correct from the expert view. 

\revise{\textbf{Identify missed elements by the model.}}
After eliminating suspicious predictive elements, the art historian discovered that the structured representation model still missed certain elements, posing challenges to semantic understanding. To address this, the historian used entity and relationship addition in the painting view to capture the missed elements. With timely expert-annotated sample recommendations, the historian accurately completed the missing elements. For example, a cross flute was added, and the corresponding relationship to a woman playing the flute is also completed (Fig. \ref{fig6}(B)).
% After eliminating suspicious predictive elements, the art historian has discovered that there are still a certain number of elements that are missed by the structured representation model, which will bring challenges to semantic understanding. 
% The art historian further used entity and relationship addition in the painting view 
% to complete the missed elements. With the help of timely expert-annotated sample recommendations, the art historian achieves accurate completion of missed elements. For example, a cross flute is added (Figure \ref{fig6}(B)).

This usage scenario demonstrates that \toolName{} can help art historians accurately and efficiently construct the structured representations of TCPs. It enables users to engage in the construction process, ensuring that the resulting structured expression achieves reliable TCP semantic understanding. 

% Involved in the construction process, the art historian felt confident about the structured representation for the semantic understanding of traditional Chinese painting.

% After accurately and efficiently constructing the structured representation using \toolName{}, the art historian achieves an in-depth semantic understanding of The Night Revels of Han Xizai based on this. Involved in the construction process, the art historian felt confident about the structured representation result.

\subsection{Expert Interviews}
To evaluate the effectiveness of \toolName{}, we interviewed six art historians, comprising two professors (AH1–AH2) with about 20 years of TCP research experience and four PhD students (AH3–AH6) focusing respectively on figure (AH3, AH5), landscape, and flower-and-bird paintings (AH4, AH6). None of the participants are co-authors of this work.

\begin{figure}[h]
	\centering
	 
 \includegraphics[width=3.5in]{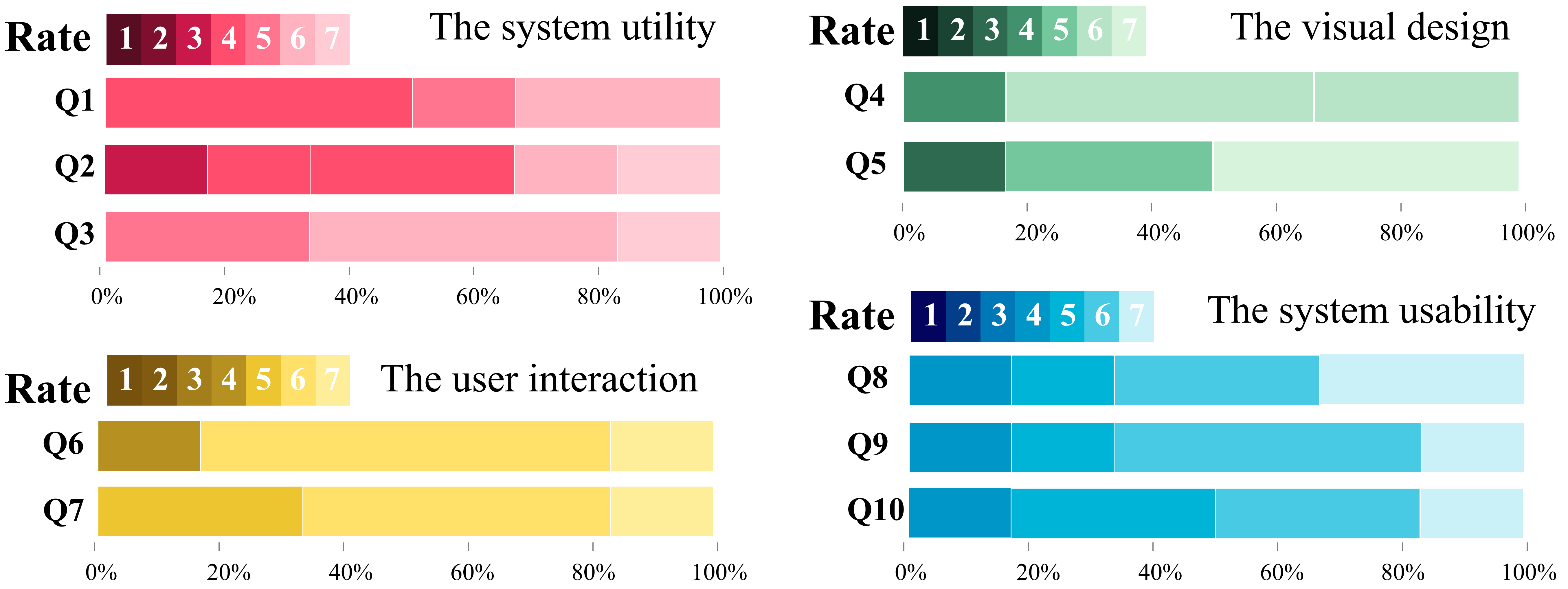}
	\caption{The feedback of the expert interviews.}
	\vspace{-2mm}
 \label{fig7}
\end{figure}

% \textbf{Procedures.} 
% For the expert interviews, 
We first introduced our motivation to enable comprehensive semantic understanding through structured representation construction with \toolName{}. Then, we showcased how to use \toolName{} and asked them to construct structured representations of their interested paintings. During the process, we recorded their comments and interactions. Subsequently, the art historians were requested to describe the findings during the construction procedure verbally. Moreover, we asked them to rate \toolName{} on a 7-point Likert scale based on the post-study questionnaire (Table \ref{table3}), focusing on 
the utility, visual design, interactions and usability of the system.
% {model performance, joint embedding visual design, interactions, and usability.} 
The post-study interview lasted about 20 minutes each. Figure \ref{fig7} shows the feedback from expert interviews. All participants provided informed consent prior to taking part in the interviews.

\textbf{Utility.}
% {Usability of the structured representation model.}} 
Most art historians ($rating_{\text{mean}}$ = 5.11, $rating_{\text{sd}}$ =1.28) offered positive feedback regarding the utility of the system in terms of the TCP-oriented knowledge extraction model($rating_{\text{mean}}$ = 4.83, $rating_{\text{sd}}$ =0.98), comparison view ($rating_{\text{mean}}$ = 5.00, $rating_{\text{sd}}$ =1.41) and expert sample recommendation ($rating_{\text{mean}}$ = 5.83, $rating_{\text{sd}}$ =0.75).
% {usability of the knowledge extraction model.} 
% \replaced{\textit{For the TCP-oriented knowledge extraction model,}}{}
% {Specifically}
First,
the art historians agreed that ``The proposed intelligent model can effectively support automatic object detection, which alleviates the repetitive and time-consuming knowledge extraction.'' 
% AH1 expressed surprise and appreciation upon discovering that the proposed model is capable of accurately identifying distinct artefacts intrinsic to ancient China, such as the pipa, cross flute, and fan.
Although other art historians also confirmed the value of the structured representation model, AH2 further said, ``The intelligent model indeed can identify those more common and easy to understand relationships, it still produces some misunderstanding of semantic events belonging to ancient China.''  
% \textit{\replaced{For comparison view,}{}
% {Intuitiveness of the visual cues.}} 
Second, most art historians are in favor of the joint embedding visualization. 
% integrated in \toolName{}. 
AH1, who used \toolName{} to construct the structured representations of a complex figure painting, commented 
% on the uncertainty visualization, 
``Being familiar with current figure painting, I find that the automatic knowledge extraction results had some misidentification, but the uncertain visualization effectiveness hinted at these errors.'' Meanwhile, four art historians (AH3-AH6) also confirmed that 
the joint embedding of predicted elements and expert-annotated samples is useful, especially in ``enhancing the credibility of automatic knowledge extraction.'' Although AH2 also expressed the advantage of the uncertainty recommendations on objects, he further emphasized the imperative for advancements in prompts related to uncertain relationships. 
% \added{\textit{For expert sample recommendation,}} 
Finally, two art historians praised the implementations of expert view in the interactive refinement component, which provides ``credible expert-annotated references during the process of constructing structured representations.''

\textbf{Visual Design.} The experts appreciated the visual design of \toolName{} ($rating_{\text{mean}}$ = 5.58, $rating_{\text{sd}}$ =1.16), especially the design of Comparison View. Most experts praised the use of scatter plots to display discrepancies between model recognition results and expert-annotated samples, and commented ``The glyph design for representing relationships effectively illustrates specific event information related to relationship triples, providing ample information for comparing and confirming relationships.'' They agreed that the design can present the distance of joint embedding vectors, and embedding vectors in respect to each identified element.

\textbf{Usability and user interaction.} The majority of the art historians provided positive feedback on user interactions ($rating_{\text{mean}}$ = 5.83, $rating_{\text{sd}}$ =0.83) and the usability ($rating_{\text{mean}}$ = 5.67, $rating_{\text{sd}}$ =1.03). Most experts agreed that the system has good usability and light workload, including ease of learning and no need for much physical or mental activity. Specifically, the comparison view retains the intuitive nature of scatter plots, making it easy to interpret without imposing additional cognitive load.
% \deleted{Notably, two art historians praised the implementations of expert view in the interactive refinement component, which can provide users with credible expert-annotated references during the process of constructing structured representations, enhancing the credible TCP understanding.} 
AH1-6 confirmed that the system's interactions are smooth and easy to use. Meanwhile, the art historians also pointed out that ``The linkage of multiple coordinated views provides more background knowledge for the interactive authentication of structured representations.'' Additionally, art historians are satisfied with the structured representation of TCPs constructed by \toolName{}, as the system ``supports a human-in-the-loop approach, allowing for the identification, confirmation, and modification of elements to ensure a reliable and semantically rich representation of the paintings.''
% {Also, AH5 expressed a strong will to recommend \toolName{} to his colleagues in his research lab.}
% Art historians are satisfied with the structured representation of TCPs constructed by VisTCP, as the system supports a human-in-the-loop approach, allowing for the identification, confirmation, and modification of elements to ensure a reliable and semantically rich representation of the paintings.

\begin{table}
\centering
\caption{The questionnaire consists of four parts: the utility (Q1-3), visual design (Q4-5), interactions (Q6-7), and the usability of the proposed \toolName{} (Q8-10).} 
\begin{tblr}{
  width = 1.14\linewidth,
  colspec = {Q[67]Q[808]Q[27]Q[27]},
  vline{2} = {-}{},
  hline{1,4,6,8,11} = {1-2}{},
  hline{2-3,5,7,9-10} = {2}{},
}
Q1       & TCP-oriented model is efficient for knowledge extraction.                             &  &  \\
Q2~ ~ ~~ & Comparison View is effective for misunderstanding and credible results prompt.                          &  &  \\
Q3~ ~ ~~ & Expert sample recommendation is useful for trustworthy results refinement. &  &  \\
Q4       & The visual design is easy to understand.              &  &  \\
Q5~ ~ ~~ & Comparison View can enable intuitive representation of the automatic knowledge extraction results.~                                                                                       &  &  \\
Q6       & The interaction is smooth.                                                   &  &  \\
Q7~ ~ ~~ & The interaction can support trustworthy structured representation confirmation and completion.                                     &  &  \\
Q8       & VisTCP is easy to learn and understand.                                                                                        &  &  \\
Q9~ ~ ~~ & VisTCP is easy to use.                              &  &
\\
Q10~ ~ ~~ & User is satisfied with constructed structured representation.                              &  &

\end{tblr}
\label{table3}
\end{table}

\section{Discussion}
% This section first outlines the potential usage scenarios
% % lessons learned from working with art historians 
% and then discusses the limitations and future work of \toolName{}.

\subsection{Potential Usage Scenario}
The construction of structured representation shows obvious benefits in TCP semantic representation. In addition, similar to the structured representation of natural images, scene graphs can support many high-level downstream tasks. On this basis, structured representations of TCPs show great potential in benefiting the following applications.

\textbf{Semantic retrieval for TCPs.} 
Scholars engaged in archaeological and art historical research often require extensive retrieval efforts to access the desired ancient paintings. 
As illustrated in the Case Study, the structured representation of TCPs can encapsulate a variety of objects and semantic events.
% present in them. 
% Consequently, they can enable art historians to explore TCPs through semantic retrieval. For instance, art historians can employ structured representations to retrieve paintings containing semantic events such as man playing the zither, facilitating in-depth analysis of the painting's content. This approach contrasts with traditional methods, which often necessitate art historians to engage in repeated retrieval and screening processes to fulfill such retrieval needs.
Consequently, they enable art historians to explore TCPs through semantic retrieval. For instance, art historians can use structured representations to find paintings depicting specific semantic events, like a man playing the zither, facilitating in-depth content analysis. This approach reduces the repeated retrieval and screening involved in traditional methods.
% This approach alleviates the need for repeated retrieval and screening required by traditional methods.

\textbf{TCP structured understanding by VLM.}
Existing literature has demonstrated that scene graph knowledge, which facilitates the structured representation of natural image content, can effectively enhance the multimodal structured understanding of large-scale visual-language models. Building upon this premise, integrating the structured representation of TCPs as knowledge-enhanced encodings into visual-language models can further augment the models' understanding of detailed structural knowledge inherent in ancient artworks, encompassing semantic relationships and attribute features.

\subsection{Limitations and Future Work}
\textbf{Advancing TCP Semantic Taxonomy.} Based on the expert study, we've developed a comprehensive TCP semantic taxonomy,  
which is specifically tailored to effectively describe the semantic content drawn in TCPs.
Collaborating with art historians further reveals a demand for a more detailed and comprehensive label system. For instance, experts note the evolving nature of objects like hats which exhibit distinct stylistic characteristics across various historical periods and geographical regions. 
In the future, we will delve deeper into the research of TCPs 
making the semantic taxonomy accessible to more complex Chinese culture.

\textbf{Incorporating diverse expert knowledge.}
In this study, we primarily furnish expert-annotated paintings as prior knowledge to facilitate the construction of credible structured representations. However, for specific understanding requirements of ancient artworks, art historians necessitate seeking additional research findings. Since the understanding of TCPs spans millennia, a plethora of documents and data records exist to aid art history analyses from various perspectives. These data, to a certain extent, assist art historians in multidimensional information understanding. In the future, we aim to further integrate multimodal data records to support more profound explorations of the semantic content in TCPs.

\textbf{Improving model performance.}  \toolName{} has developed an intelligent knowledge extraction model tailored for TCPs. However, the specialized nature of TCPs and the sparsity of annotated samples pose challenges to the current system's utility, especially uncertainty prompts and expert-annotated sample recommendations for relationship inference. Therefore, the system offers interactions and model iteration optimization based on expert feedback. We anticipate that through continual iterative improvements, the system's performance in constructing structured representations of TCPs will be continuously optimized.

\textbf{Generalizability.} In this work, we primarily collaborated with art historians to develop a human-machine collaborative system to facilitate the structured representation construction of TCP. Although this system is specifically designed for TCPs, the proposed framework can be extended to other domain-specific image types, particularly other art and natural images. We learned that the accuracy of knowledge extraction by the automatic intelligent method is far from usable. The scene graph of natural images constructed by our interactive system can fully guarantee the quality of the constructed graph, enabling its downstream roles.
Furthermore, by appropriately defining domain-adapted semantic taxonomies and intelligent models, the system's comparison view and expert sample recommendation strategy can easily handle various professional image data, such as medical images, autonomous driving, and industrial inspection. For example, the comparison view based on medical images can help doctors quickly identify abnormal results, and expert sample recommendations can aid doctors in achieving reliable disease diagnoses based on historical experience.

\section{Conclusion}
In this paper, we present \toolName{}, a visualization framework that facilitates art historians to achieve structured representations of TCPs in an efficient and trustworthy way. The framework integrates the proposed TCP-oriented knowledge extraction model and prior expert knowledge, to support interactions between art historians and machines. We develop an interface with a joint embedding visualization to support art historians’ understanding of structured representation quality and to highlight uncertainties for interactive refinement.
The effectiveness and practical utility of \toolName{} are demonstrated through a case study, a usage scenario, and expert interviews. The results reveal that the structured representation can effectively enable comprehensive TCP understanding and the framework interface offers art historians an efficient way to interact with intelligent models for improving structured representation. 

\section{Acknowledgements}
% Acknowledgements
% This work was supported in part by the National Natural Science Foundation of China (Nos. 62277013 and 62177040), the
% National Statistical Science Research Project (No. 2023LZ035),
% the Key R\&D ‘Jianbin’ Tackling Plan Programme in Zhejiang
% Province, China (No. 2023C01119), the Public Welfare Plan Research Project of Zhejiang Provincial Science and Technology
% Department (Nos. LTGG23H260003 and LGF22F020034), Open
% Project Programme of the State Key Laboratory of CAD\&CG (No.
% A2301) and Zhejiang Provincial Natural Science Foundation of
% China (No. LTGG24F020006).
This work was supported in part by the National Natural Science Foundation of China (Nos.~62277013 and~62177040), 
the National Social Science Fund Major Project (No.~24\&ZD075), 
the National Statistical Science Research Project (No.~2023LZ035), 
the Key R\&D 'Jianbin' Tackling Plan Programme in Zhejiang Province, China (No.~2023C01119), 
the Public Welfare Plan Research Project of Zhejiang Provincial Science and Technology Department (Nos.~LTGG23H260003, LGF22F020034 and~LZ22F020015), 
the Open Project Programme of the State Key Laboratory of CAD\&CG (No.~A2301), 
and the Zhejiang Provincial Natural Science Foundation of China (No.~LTGG24F020006).

\bibliographystyle{IEEEtran}

\bibliography{template}
% \bibliographystyle{IEEEtran}
% \bibliography{haha}

%% ^^^^^   FOR IEEE VIS, EVERYTHING HERE MAY BE INCLUDED IN THE    ^^^^^ %%
%% 2-PAGE ALLOTMENT FOR REFERENCES, FIGURE CREDITS, AND ACKNOWLEDGEMENTS %%

% \appendix % You can use the `hideappendix` class option to skip everything after \appendix

% \begin{IEEEbiography}[{\includegraphics[width=1in,height=1.25in,clip,keepaspectratio]{yeli}}]{Li Ye}
% hahahahahalalalala
% \end{IEEEbiography}

% \begin{IEEEbiography}[{\includegraphics[width=1in,height=1.25in,clip,keepaspectratio]{yeli}}]{Li Ye}
% hahahahahalalalala
% \end{IEEEbiography}

% \begin{IEEEbiography}[{\includegraphics[width=1in,height=1.25in,clip,keepaspectratio]{yeli}}]{Li Ye}
% hahahahahalalalala
% \end{IEEEbiography}

% \begin{IEEEbiography}[{\includegraphics[width=1in,height=1.25in,clip,keepaspectratio]{yeli}}]{Li Ye}
% hahahahahalalalala
% \end{IEEEbiography}

\end{document}